\documentstyle[12pt]{article}

\makeatletter \@addtoreset{equation}{section}

\newlength{\defaultparindent}
\newenvironment{HTML Preformatted}{}{}
\newenvironment{Body Text}{}{}
\newenvironment{Default Paragraph Font}{}{}
\input{tcilatex}
\begin{document}

\title{{\bf {\bf Explicit derivation of a Central extended Hyper-Kahler Metric}.}}
\author{M. HSSAINI$^{1}$, B. MAROUFI$^{1}$, M. KESSABI$^{1}$, M.B.SEDRA $^{1,\,2}$ \thanks{
Corresponding author: sedra@ictp.trieste.it} \\
$^{1}${\small \ \ Laboratoire de Physique des Hautes Energies,
Facult\'{e} des Sciences,}\\ {\small \ \ \ \ P.O. Box 1014, Ibn.
Battouta, Rabat, Morocco,}\\ {\small \ \ \ \ }$^{1,\,2}${\small \
Laboratoire de Physique Th\'{e}orique et Appliqu\'{e}e (LPTA),}\\
{\small \ \ \ \ Facult\'{e} des Sciences, D\'{e}partement de
Physique, B.P. 133, K\'{e}nitra, Morocco}{\tiny .}} \maketitle

\begin{abstract}
This work consists in applying the analysis of integrable models to study
the problem of Hyper-Kahler metrics building. In this context, we use the
harmonic superspace language applied to D=2 N=4 SU(2) Liouville self
interacting model and derive an explicit central extended Hyper-Kahler
metric as well as the induced scalar potential.
\end{abstract}

\bigskip

\newpage

\section{Introduction:}

Hyper-kahler metrics building program is an important question of
Hyperkahler geometry in two and four dimensions and is solved in a
nice way in the harmonic superspace [1-3]. More recently it was
found that moduli spaces arising in topological field theories and
other areas of mathematical physics possess Hyper-kahler
structures, a notable example being the moduli space of BPS
magnetic monopoles [4].

Similar structures are also encountered in superstring theory, and
in many other moduli problems of physical interest in quantum
gravity [5-7]. In spite of these achievements, many things remain
to be done. The classification of all complete, regular,
Hyperkahler manifolds remains an open question to this date, and
even for some known examples, the explicit form of the metric has
been difficult or impossible to determine so far. Recall that
there are a few examples scattered in the literature that have
been solved exactly.

Two of these examples are provided by the Eguchi-Hanson and the
Taub-Nut metrics exhibiting both $U(2)=SU(2)\times U(1)$
isometries. Another example which is derived explicitly is the
Hyperkahler metric associated to the harmonic superspace Toda
(Liouville) -like self- interaction and which can be useful in the
sense that the particular Hyperkahler geometrical structure that
it is expected to describe, is connected to integrable models via
the Toda-like self- interaction [3,8,9]. The explicit construction
of a new series of metrics will certainly require a better and
more systematic understanding of the unusual features that arise
in the continual limit of the Toda field equations. In the present
work, we compute explicitly the metric associated to D=2 N=4 SU(2)
Liouville Hyper-Kahler action in the harmonic superspace in the
presence of central charges. We review in section 2 some general
properties of the Hyper-Kahler metric building program. Focusing
to obtain the central extended bosonic action, we develop in
section 3 some integrability techniques to derive the auxiliary
fields $F^{--}$, $G^{--}$ and the Lagrange field $ \Delta$, and
present a method for their exact solvability. In section 4 we give
the details concerning the derivation of the central extended
metric as well as the induced scalar potential. The section 5 is
devoted to our conclusion and discussion.

\section{Generalities on Hyper-Kahler metrics building program in D=2 N=4
HS}

We recall in this section some general results of the Hyper-kahler
metric building from harmonic superspace. The subject of
Hyper-kahler metrics building is an interesting problem of
Hyper-kahler geometry that can be solved in a nice way in harmonic
superspace if one knows how to solve the following non-linear
differential equations on the sphere S$^{2}$[2].

\begin{eqnarray}
\partial^{++}q^{+}-\partial^{++}\left[ \frac{\partial V^{4+}}{\partial
\left( \partial ^{++}\overline{q}^{+}\right) }\right] +\frac{\partial V^{4+}%
} {\partial \overline{q}^{+}}&=&0  \nonumber \\
\partial^{++}q^{+}+\partial^{++}\left[ \frac{\partial V^{4+}}{\partial
\left( \partial^{++}q^{+}\right) }\right] -\frac{\partial V^{4+}}{\partial
q^{+}}&=&0
\end{eqnarray}

where $q^{+}=q^{+}\left( z,\overline{z},u^{\pm }\right) $ and its conjugates
$\overline{q}^{+}=\overline{q}^{+}\left( z,\overline{z},u^{\pm }\right) $
globally defined on $C\times S^{2} \left( S^{2}\approx SU(2)\times
U(1)\right)$ are complex fields parameterized by the local complex
coordinates $\left( z,\overline{z}\right) $ and the harmonic variables $%
u^{\pm }$ with . The symbol $\partial ^{++}=u^{+i}\frac{\partial }{\partial
u^{-i}}$ stands for the so-called harmonic derivative and $%
V^{4+}=V^{4+}\left( q,u\right) $ is an interacting potential depending in
general on $q^{+},\overline{q}^{+}$; their derivatives and on the $u^{\pm }$
$^{\prime }s$ . Note by the way that $q^{+},\overline{q}^{+}$ may be
expanded into an infinite series in powers of harmonic variables (for the
bosonic part) preserving then the total U(1) charge in each term of the
expansion as shown here below:

\begin{equation}
q^{+}\left( z,\overline{z},u\right) =u_{i}^{+}f^{i}\left( z,\overline{z}%
\right) +u_{i}^{+}u_{j}^{+}u_{k}^{-}f^{\left( ijk\right) }\left( z,\overline{%
z}\right) +...
\end{equation}
Note also that (2.1), which fixes the u-dependence of the $q^{+}$ $^{\prime
}s$ , is in fact the pure bosonic projection of a two- dimensional N=4
supersymmetric HS superfield equation of motion. The remaining equations
carry the spinor contributions and are shown to describe, among others, the
space-time dynamics of the physical degrees of freedom namely $f^{i}\left( z,%
\overline{z}\right) $, $\overline{f}^{i}\left( z,\overline{z}\right) ,$
i=1,2 and their D=2, N=4 supersymmetric partners. By using the
How-Stelle-Townsend (HST) realization of D=2, N=4 hypermultiplet $\left(
O^{4},\left( \frac{1}{2}\right) ^{4}\right) $[10], Eqs.(2.1) read as

\begin{equation}
\partial ^{++^{2}}\omega -\partial ^{++}\left[ \frac{\partial H^{4+}}{%
\partial (\partial ^{++}\omega )}\right] \,+\frac{\partial H^{4+}}{\partial
\omega }=0
\end{equation}
where $\omega =\omega (z,\bar{z},u)$ is a real field with zero U(1) charge
defined on $C\times S^{2}$ and whose leading terms of its harmonic expansion
given by

\begin{equation}
\omega \left( z,\overline{z},u\right) =u_{i}^{+}u_{j}^{-}f^{ij}\left( z,%
\overline{z}\right) +u_{i}^{+}u_{j}^{+}u_{k}^{-}u_{l}^{-}g^{ijkl}\left( z,%
\overline{z}\right) +...
\end{equation}
Similar as in (2.1), the interacting potential $H^{4+}$ depends in general
on $\omega $ , its derivatives and the harmonics. Note the important
observation of [11], that one can always pass from the $q^{+}$
hypermultiplet to the $\omega $ hypermultiplet via a duality transformation
[12] by making a change of variables. In the remarkable case where the
potentials $H^{4+}$ and $V^{4+}$ do not depend on the derivatives of the
fields $q^{+}$ and $\omega $ (2.1,3) reduce then to

\begin{equation}
\partial ^{++}q^{+}+\frac{\partial V^{4+}}{\partial \overline{q}^{+}}=0
\end{equation}

\begin{equation}
\partial ^{++^{2}}\omega +\frac{\partial H^{4+}}{\partial \omega }=0
\end{equation}
Since the solutions of these equations depend naturally on the potentials $%
V^{4+}$ and $H^{4+}$, finding these solutions is not an easy task. There are
only few examples scattered in the literature that have been solved exactly
[2,9,13]. Let us review in what follows some them. The first example is the
Taub-Nut (T.N) metric of D=4 Euclidean gravity. Its Potential $V^{4+}$ reads
as [2]

\begin{equation}
V^{4+}\left( q^{+},\overline{q}^{+}\right) =\frac{\lambda }{2}\left( q^{+}%
\overline{q}^{+}\right) ^{2},
\end{equation}
where $\lambda $ is a real coupling constant. According to this potential,
the equation of motion is

\begin{equation}
\partial ^{++}q^{+}+\lambda q^{+}\left( \overline{q}^{+}\right) q^{+}=0
\end{equation}
whose solution reads as

\begin{equation}
q^{+}(z,\bar{z},u)=u_{i}^{+}\,f^{i}(z,\bar{z})\,\exp (-\lambda
u_{k}^{+}u_{l}^{-}f^{(k}f^{l)}).
\end{equation}
Note that the knowledge of this solution is an important step towards
identifying the metric of the manifold parameterized by $f^{i}\left( z,%
\overline{z}\right) $ and $\overline{f}^{i}\left( z,\overline{z}\right) $ of
the D=2 N=4 supersymmetric non linear Taub-Nut $\sigma $ -model. The latter
possesses an action whose bosonic part reads as

\begin{equation}
S_{B}^{TN}=-\frac{1}{2}\int dzd\overline{z}\left( g_{ij}\partial
_{z}f^{i}\partial _{\overline{z}}f^{j}+\overline{g}^{ij}\partial _{z}%
\overline{f}_{i}\partial _{\overline{z}}\overline{f}_{j}+2h_{j}^{i}\partial
_{z}f^{j}\partial _{\overline{z}}\overline{f}_{i}\right) .
\end{equation}
and the metric is known to be[14]

\begin{equation}
\left.
\begin{array}{c}
g_{ij}=\frac{\lambda \left( 2+\lambda f\overline{f}\right) }{2\left(
1+\lambda f\overline{f}\right) }\overline{f}_{i}\overline{f}_{j},\text{ }%
\overline{g}^{ij}=\frac{\lambda \left( 2+\lambda f\overline{f}\right) }{%
2\left( 1+\lambda f\overline{f}\right) }f_{i}f_{j}, \\
h_{j}^{i}=\delta _{j}^{i}\left( 1+\lambda f\overline{f}\right) -\frac{%
\lambda \left( 2+\lambda f\overline{f}\right) }{2\left( 1+\lambda f\overline{%
f}\right) }f_{i}\overline{f}_{j}\text{ }and\text{ }f\overline{f}\equiv f^{i}%
\overline{f}_{i}.
\end{array}
\right.
\end{equation}
In the standard Taub-Nut form [15], we have

\begin{equation}
ds^{2}=\frac{r+M}{2\left( r-M\right) }dr^{2}+\frac{1}{2}\left(
r^{2}-M^{2}\right) \left( d\theta ^{2}+\sin ^{2}\theta d\varphi ^{2}\right)
+2M^{2}\left( \frac{r-M}{r+M}\right) \left( d\psi +\cos \theta d\varphi
\right) ^{2}
\end{equation}
once the following change of variables [14]

\begin{equation}
\left.
\begin{array}{c}
f^{1}=\sqrt{2M\left( r-M\right) }\cos \frac{\theta }{2}\exp \frac{i}{2}%
\left( \psi +\varphi \right) , \\
f^{2}=\sqrt{2M\left( r-M\right) }\sin \frac{\theta }{2}\exp \frac{i}{2}%
\left( \psi -\varphi \right) ,
\end{array}
\right.
\end{equation}
is performed with $f\overline{f}=2M\left( r-M\right) ,$ $r > M \equiv \frac{1%
}{2\sqrt{\lambda }}.$

The second example we consider is the Eguchi-Hanson (E.H) model.
This has also been solved exactly and corresponds to the following
potential:

\begin{equation}
H^{4+}\left( \omega \right) =\left[ u_{i}^{+}u_{j}^{+}\xi ^{\left( ij\right)
}\right] ^{2}\times \omega ^{2}
\end{equation}
where is an SU(2) real constant triplet. Thus, unlike the T.N action, the
E.H one contains explicit harmonics. More details are exposed in [2].

Recently, a new integrable model has been proposed [9]. This model was
obtained by focussing on (2.5b) and looking for potentials leading to exact
solutions of this equation. The method used in this issue consist in
suggesting new plausible integrable equations by proceeding in formal
analogy with the known integrable two dimensional non-linear differential
equations, especially the Liouville equation and its Toda generalizations
[16].

The important result in this sense was the proposition of the following
potential:

\begin{equation}
H^{4+}(\omega ,u)=\left( \frac{\xi ^{++}}{\lambda }\right) ^{2}\exp
(2\lambda \omega ),
\end{equation}
which leads, to the following non linear differential equation of motion:

\begin{equation}
\lambda \left( \partial ^{++}\right) ^{2}\omega -\xi ^{++^{2}}e^{2\lambda
\omega }=0.
\end{equation}
Using the formal analogy with the SU(2) Liouville equation, we showed that
(2.15) is integrable. The explicit solution of this non-linear differential
equation is [9]

\begin{equation}
\xi ^{++}e^{\lambda \omega }=\frac{u_{i}^{+}u_{j}^{+}f^{ij}\left( z,%
\overline{z}\right) }{1-u_{k}^{+}u_{l}^{-}f^{kl}\left( z,\overline{z}\right)
}.
\end{equation}
Furthermore, the origin of the integrability in (2.15) is shown to deal with
the existence of a symmetry (conformal symmetry) generated by the following
conserved current

\begin{equation}
T^{4+}=\partial ^{++}{}^{2}-\frac{1}{\lambda }\partial ^{++}{}^{2}\omega ,
\end{equation}
with $\partial ^{++}T^{4+}=0$ . Note that other examples of Hyper-Kahler
potentials are known in the literature [13].

\section{Central extension of the bosonic action and integrability:}

\begin{Body Text}
Focusing in this section on the above SU(2) Liouville model, to derive the
corresponding extended Hyper-Kahler metric associated to the proposed
potential (2.14) in the presence of central charges. The general procedure
to get the component form of the bosonic non-linear sigma-model using the
harmonic superspace approach consists in applying the method presented in
[2].One starts by writing the action describing the general coupling of the
analytic superfield $\Omega $

we are interested in and deriving the corresponding equation of motion. For
this action, corresponding to some Hyper-Kahler manifold, one only has to
expand the equations of motion in Grassman variables $\theta $ and ignore
all the fermionic field components. Then one has to solve the kinematical
differential equations on the sphere S$^{2}$ for the auxiliary field
components a fact which leads to eliminate the infinite tower of them in the
harmonic expansion of the hypermultiplet HSS superfields. Substituting the
solution into the original action in HSS and integrating over all the
harmonic variables and the anti-commuting coordinates yields the required
component form of the action form which the Hyper-Kahler metric can be
extracted.

Let us first recall that the harmonic superspace (HSS) is parameterized by
the super-coordinates $Z^{M}=\left( Z_{A}^{M},\theta _{r}^{-},\overline{%
\theta }_{r}^{-}\right) ,$ where $Z_{A}^{M}=\left( Z,\overline{Z},\theta
_{r}^{+},\overline{\theta }_{r}^{+},u^{\pm }\right) $ are the super
coordinates of the so-called analytic subspace in which D=2 N=4
supersymmetric theories are formulated. The integral measure of the harmonic
superspace is given in the $Z_{A}^{M}$ basis by $d^{2}Zd^{4}\theta ^{+}du.$
The matter superfield $\left( O^{4}\left( 1/2\right) ^{4}\right) $ is
realized by two dual analytic superfields $Q^{+}=Q^{+}\left( Z,\overline{Z}%
,\theta ^{+},\overline{\theta }^{+},u\right) $ and $\Omega =\Omega (z,%
\overline{z},\theta ^{+},\overline{\theta }^{+},u),$ whose leading bosonic
fields are respectively given by q+ and $\omega $ Eqs(2.2, 4). The model we
are interested in and which describes the coupling of the analytic
superfield $\Omega ,$ is given by the following action [9]
\end{Body Text}

\begin{equation}
S[\Omega ]=\int d^{2}zd^{4}\theta ^{+}du\left\{ \frac{1}{2}\left(
D^{++}\Omega \right) ^{2}+\frac{1}{2}\left( \frac{\xi ^{++}}{\lambda }%
\right) ^{2}e^{2\lambda \Omega }\right\}
\end{equation}
where $\lambda $ is the coupling constant of the model and $\xi
^{++}=u_{i}^{+}u_{j}^{+}\xi ^{ij}$ a constant isotriplet similar to that
appearing in the E.H model and where$D^{++}$ is the harmonic derivative
whose central extension is [17]

\begin{equation}
D^{++}=\partial ^{++}-2\overline{\theta }_{+}^{+}\theta _{+}^{+}\partial
_{--}-2\overline{\theta }_{-}^{+}\theta _{-}^{+}\partial _{++}+\theta
_{+}^{+}\theta _{-}^{+}\overline{Z}-\overline{\theta }_{-}^{+}\overline{%
\theta }_{+}^{+}Z
\end{equation}
where the central charges $(Z,\bar{Z})$ are generators which belongs to the
Cartan subgroup of a given Lie group. The HSS equation of motion for the
analytic superfield $\Omega $ reads:

\begin{equation}
\lambda D^{++^{2}}\Omega -\xi ^{++^{2}}e^{2\lambda \Omega }=0
\end{equation}
where $\Omega $ can by expanded in a series of $\theta _{r}^{+}$ and $\bar{%
\theta}_{r}^{+},r=\pm $ as

\begin{equation}
\left.
\begin{array}{c}
\Omega =\omega +\left[ \theta _{-}^{+}\theta _{+}^{+}F^{--}+\bar{\theta}%
_{+}^{+}\bar{\theta}_{-}^{+}\bar{F}^{--}\right] +\left[ \bar{\theta}%
_{-}^{+}\theta _{+}^{+}G^{--}+\bar{\theta}_{+}^{+}\theta _{-}^{+}\bar{G}%
^{--}\right] \\
+\left[ \bar{\theta}_{-}^{+}\theta _{-}^{+}B_{++}^{--}+\bar{\theta}%
_{+}^{+}\theta _{+}^{+}B_{--}^{--}\right] +\left[ \bar{\theta}_{-}^{+}\theta
_{-}^{+}\bar{\theta}_{+}^{+}\theta _{+}^{+}\Delta ^{-4}\right]
\end{array}
\right.
\end{equation}
In the presence of central charges, some of the kinematical equations of
motion in the $\left( Z_{A},u\right) $ space get modified

\begin{equation}
\lambda \partial ^{++}{}^{2}\omega -\xi ^{++}{}^{2}e^{2\lambda \omega }=0
\end{equation}

\begin{equation}
\partial ^{++}{}^{2}F^{--}-2\xi ^{++}{}^{2}F^{--}e^{2\lambda \omega }-2\bar{Z%
}\partial ^{++}\omega =0
\end{equation}

\begin{equation}
\partial ^{++}{}^{2}\bar{F}^{--}-2\xi ^{++}{}^{2}\bar{F}^{--}e^{2\lambda
\omega }+2Z\partial ^{++}\omega =0
\end{equation}

\begin{equation}
\partial ^{++}{}^{2}G^{--}-2\xi ^{++}{}^{2}G^{--}e^{2\lambda \omega }=0
\end{equation}

\begin{equation}
\partial ^{++}{}^{2}\bar{G}^{--}-2\xi ^{++}{}^{2}\bar{G}^{--}e^{2\lambda
\omega }=0
\end{equation}

\begin{equation}
\partial ^{++}{}^{2}B_{++}^{--}-2\xi ^{++}{}^{2}B_{++}^{--}e^{2\lambda
\omega }=4\partial ^{++}\partial _{++}\omega
\end{equation}

\begin{equation}
\partial ^{++}{}^{2}B_{--}^{--}-2\xi ^{++}{}^{2}B_{--}^{--}e^{2\lambda
\omega }=4\partial ^{++}\partial _{--}\omega
\end{equation}

\begin{equation}
\left.
\begin{array}{c}
\partial ^{++}{}^{2}\Delta ^{-4}-4\partial ^{++}\partial
_{--}\,B_{++}^{--}-4\partial ^{++}\partial _{++}\,B_{--}^{--}+8\partial
_{--}\partial _{++}\,\omega \\
-2\bar{Z}\partial ^{++}\bar{F}^{--}+2Z\partial ^{++}F^{--}-2\left| Z\right|
^{2}\omega \\
=\,2\,\xi ^{++}{}^{2}e^{2\lambda \omega }\left\{ \Delta ^{-4}+2\lambda
\,\left( F^{--}\bar{F}^{--}-G^{--}\bar{G}^{--}+B_{++}^{--}B_{--}^{--}\right)
\right\}
\end{array}
\right.
\end{equation}
The Liouville like equation of motion (3.5) is a constraint equation fixing
the dependence of $\omega $ in terms of the physical bosonic fields $%
f^{\left( ij\right) }$ of the D=2 N=4 hypermultiplet. As it is already
discussed in [3,9]; the Knowledge of the explicit solution of this
non-linear differential equation is a crucial step in this program. The
second set of relations (3.6-8) describes the equations of motion of the
auxiliary fields $F^{--}$ and $G^{--}$ and of canonical dimension one. (3.9)
gives the equation of motion of the Lagrange field of canonical dimension
two in terms of and the other auxiliary fields. Note that the principal
contributions of the central charges appear only on the auxiliary fields $%
F^{--},$ $\overline{F}^{--}$ and the Lagrange field $\Delta ^{-4}.$
Furthermore, the fields $F^{--}$ and $G^{--}$ are shown to behave in a
similar way in the absence of central charges. To solve these equations of
motion, one starts first by solving the Liouville-like equation of motion
(3.5) whose solution [16]; originated from integrability and conformal
symmetry in two dimensions is given in the HS language by [9]

\begin{equation}
\lambda \partial ^{++}\omega =\xi ^{++}e^{\lambda \omega }=\frac{f^{++}}{1-f}%
\text{ ,}
\end{equation}
with

\begin{equation}
\left.
\begin{array}{c}
f=u_{(i}^{+}u_{j)}^{-}\,f^{(ij)}+f^{0} \\
f^{++}=u_{(i}^{+}u_{j)}^{+}\,f^{(ij)}
\end{array}
\right.
\end{equation}
After integrating over the Grassman variables in the action (3.1) one finds
that the bosonic action reduces to

\begin{equation}
\left.
\begin{array}{c}
S=\int d^{2}zdu\left\{ \left[ \partial ^{++}\omega \partial ^{++}\Delta
^{-4}+\frac{1}{\lambda }\xi ^{++^{2}}e^{2\lambda \omega }\Delta ^{-4}\right]
\right. \\
+\left[ \partial ^{++}F^{--}\partial ^{++}\overline{F}^{--}+2\xi
^{++^{2}}e^{2\lambda \omega }F^{--}\overline{F}^{--}\right] \\
-\left[ \partial ^{++}G^{--}\partial ^{++}\overline{G}^{--}+2\xi
^{++^{2}}e^{2\lambda \omega }G^{--}\overline{G}^{--}\right] \\
+\left[ \partial ^{++}B_{--}^{--}\partial ^{++}B_{++}^{--}-2\partial
^{++}\omega \left( \partial _{--}B_{++}^{--}+\partial
_{++}B_{--}^{--}\right) \right] \\
+\left[ 4\partial _{--}\omega \partial _{++}\omega -2\left( \partial
_{++}\omega \partial ^{++}B_{--}^{--}+\partial _{--}\omega \partial
^{++}B_{++}^{--}\right) +2\xi ^{++^{2}}e^{2\lambda \omega
}B_{++}^{--}B_{--}^{--}\right] \\
-\left. \overline{Z}\left( \partial ^{++}\omega \overline{F}^{--}+\partial
^{++}\overline{F}^{--}\omega \right) +Z\left( \partial ^{++}\omega
F^{--}+\partial ^{++}F^{--}\omega \right) -\left| Z\right| ^{2}\omega
^{2}\right\}
\end{array}
\right.
\end{equation}
Using the equation of motion for $\omega $ (3.5), one shows that the
following harmonic integrals vanish

\begin{equation}
\int du\left[ \partial ^{++}\omega \partial ^{++}\Delta ^{-4}+\frac{1}{%
\lambda }\xi ^{++^{2}}e^{2\lambda \omega }\Delta ^{-4}\right] =0,
\end{equation}

\begin{equation}
\int du\left[ \partial ^{++}G^{--}\partial ^{++}\overline{G}^{--}+2\xi
^{++^{2}}e^{2\lambda \omega }G^{--}\overline{G}^{--}\right] =0,
\end{equation}

\begin{equation}
\int du\overline{Z}\left( \partial ^{++}\omega \overline{F}^{--}+\partial
^{++}\overline{F}^{--}\omega \right) =0,
\end{equation}

\begin{equation}
\int duZ\left( \partial ^{++}\omega F^{--}+\partial ^{++}F^{--}\omega
\right) =0.
\end{equation}

These vanishing integrals serve to eliminate the auxiliary fields $G$ and $%
\Delta $ . The resulting bosonic action is then

\begin{equation}
\left.
\begin{array}{c}
S=\int d^{2}zdu\left( 4\partial _{--}\omega \partial _{++}\omega -2\partial
^{++}\omega \left( \partial _{--}B_{++}^{--}+\partial _{++}B_{--}^{--}+%
\overline{Z}\overline{F}^{--}\right) \right. \\
+\left. 2\partial _{++}\omega \partial ^{++}B_{--}^{--}-2\partial
_{--}\omega \partial ^{++}B_{++}^{--}-\left| Z\right| ^{2}\omega ^{2}\right)
\end{array}
\right.
\end{equation}
Note that at $\left( Z,\overline{Z}\right) =0$, the auxiliary field F now
contribute too unlike that in ref[3]. To obtain a purely bosonic theory; one
have to reduce much more this action. To do this; one needs to solve the non
linear differential equation for the auxiliary fields F and B. Using some
algebraic manipulations based on the knowledge of the Liouville-like
solution (3.10) and requiring the consistency we propose the following
solutions for F and B (3.6,8)

\begin{equation}
\left.
\begin{array}{c}
F^{--}=\frac{1}{\xi ^{++}}\left\{ -\frac{\overline{Z}}{\lambda }e^{-\lambda
\omega }+\overline{\alpha }e^{\lambda \omega }\right\} \\
\overline{F}^{--}=\frac{1}{\xi ^{++}}\left\{ +\frac{Z}{\lambda }e^{-\lambda
\omega }-\alpha e^{\lambda \omega }\right\}
\end{array}
\right.
\end{equation}
and

\begin{equation}
B_{rr}^{--}=\frac{1}{\xi ^{++}}\left\{ -2\partial _{rr}\omega \,e^{-\lambda
\omega }+\eta _{rr}\,e^{\lambda \omega }\right\} ,\,\,\,\,\,\,\,r=\pm
\end{equation}
where $\alpha \,\,\,and\,\,\eta $ are two arbitrary constants which satisfy $%
\bar{\bar{\alpha}}=-\alpha $ and $\partial ^{++}\eta _{rr}=0$ .
(3.18) are obtained by proceeding by steps. The first step
consists in finding an homogeneous solution (for $Z=\bar{Z}=0$ )
and then propose a particular
solution leading then to the general form (3.18). The explicit solution for $%
B_{rr}^{--}$ contains also an homogeneous solution ($\partial ^{++}\partial
_{--}\omega =0$ )

\begin{equation}
B_{rr}^{--}=\frac{\eta _{rr}}{\xi ^{++}}e^{\lambda \omega },
\end{equation}
and a particular one

\begin{equation}
B_{rr}^{--}=\frac{-2}{\xi ^{++}}\partial _{rr}\omega \,e^{-\lambda \omega }
\end{equation}
Its worth pointing out that the particular solution for F and B are simply
obtained thanks to the Liouville (3.5) just by assuming that $\partial
^{++^{2}}F^{--}=0$ and $\partial ^{++^{2}}B_{rr}^{--}=0$ . Furthermore note
that another particular solution of $B_{rr}^{--}$ was proposed in [3] namely

\begin{equation}
B_{rr}^{--}=\xi ^{--}\partial _{rr}\omega
\end{equation}
where $\xi ^{--}=u_{(i}^{-}u_{j)}^{-}\xi ^{(ij)}$ is an SU(2) triplet
constant required to satisfy $\partial ^{++}\xi ^{--}=2$ . Although they
appear to be different, the two particular solutions (3.21, 3.22) share
naturally a crucial property namely

\begin{equation}
\partial ^{++}\xi ^{--}=\partial ^{++}(-2\frac{e^{-\lambda \omega }}{\xi
^{++}})=2.
\end{equation}
On the other hand; setting $Z=\bar{Z}=0$ in (3.6), one can derive a solution
for the auxiliary fields $G^{--}\,and\,\bar{G}^{--}$ induced from the
striking resemblance between the equations for F and G. One have

\begin{equation}
\left.
\begin{tabular}{l}
$G^{--}=\frac{\bar{\alpha}}{\xi ^{++}}e^{\lambda \omega }$ \\
$\bar{G}^{--}=-\frac{\alpha }{\xi ^{++}}e^{\lambda \omega }$%
\end{tabular}
\right.
\end{equation}
Using the previous results; note also that when $Z=\bar{Z}=0$ ; the Lagrange
field $\Delta ^{-4}$ is shown to satisfy

\begin{equation}
\left.
\begin{array}{c}
\partial ^{++2}\Delta ^{-4}-4e^{2\lambda \omega }\left[ 4\lambda (\eta
_{++}\partial _{--}\omega +\eta _{--}\partial _{++}\omega )+\partial
_{--}\eta _{++}+\partial _{++}\eta _{--}\right] +8\partial _{--}\partial
_{++}\omega \\
=2\xi ^{++2}\Delta ^{-4}e^{2\lambda \omega }+4\lambda (\eta _{++}\eta
_{--}e^{4\lambda \omega }+4\partial _{++}\omega \partial _{--}\omega ).
\end{array}
\right.
\end{equation}
whose solution ; upon setting for simplicity $\eta _{rr}=0$ ; is

\begin{equation}
\Delta ^{-4}=\frac{1}{\xi ^{++2}}\left( \bar{\alpha}e^{\lambda \omega
}+4e^{-2\lambda \omega }\left[ \partial _{--}\partial _{++}\omega -2\lambda
\partial _{++}\omega \partial _{--}\omega \right] \right) .
\end{equation}
To summarize, we have presented in this section the integrability mechanism
applied to the problem of Hyper-Kahler metrics building in the case of the
central extension of D=2 N=4 SU(2) Liouville self interaction. To accomplish
this program we have showed that the resulting system of non linear
differential equations is integrable and derived the explicit solutions for
the auxiliary fields F, B, G and $\Delta $ . Note that only F and B which
are needed in computing the pure bosonic action (3.17). The latter is shown
to correspond to the following form

\begin{equation}
\left.
\begin{array}{c}
S=\int d^{2}zdu\left\{ -4\partial _{--}\omega \partial _{++}\omega +\frac{8}{%
\lambda }\partial _{--}\partial _{++}\omega +e^{2\lambda \omega }\left[
-4\eta _{++}\partial _{--}\omega \right. \right. \\
\left. -\frac{2}{\lambda }\left( \partial _{--}\eta _{++}+\partial _{++}\eta
_{--}-\alpha \overline{Z}\right) \right] \left. -\left| Z\right| ^{2}\left(
\omega ^{2}+\frac{2}{\lambda ^{2}}\right) \right\}
\end{array}
\right.
\end{equation}
giving then the dependence of the action only on the bosonic degrees of
freedom $f=u_{i}^{+}u_{j}^{-}f^{ij}$. What remains to be done now is to use
the equation of motion (3.10) for $\omega $ and integrate over the harmonic
variables $u^{\pm }$ to derive the purely bosonic action, from which one can
easily identify the metric associated to the D=2 N=4 SU(2) Liouville self
interaction.

\section{4-The building and the structure of the extended Hyper-Kahler
metric:}

Starting from the action (3.27) and using the convenient
parameterization [9]

\begin{equation}
\xi ^{++}e^{\lambda \omega }=\frac{f^{++}}{1-f}
\end{equation}
we have

\begin{equation}
\partial _{rr}\omega =\frac{1}{\lambda }\left[ \frac{\partial _{rr}f^{++}}{%
f^{++}}+\frac{\partial _{rr}f}{1-f}\right] .
\end{equation}
Furthermore one way to write , which appears in the last term of the action
(3.27); in terms of the bosonic degrees of freedom f, is to consider for
simplicity the following ''approximation''

\begin{equation}
\omega ^{2}\approx \frac{e^{\lambda \omega }+e^{-\lambda \omega }-2}{\lambda
^{2}}
\end{equation}
which leads to set

\begin{equation}
\omega ^{2}\approx \frac{1}{\lambda ^{2}}\left[ \frac{f^{++}}{\xi
^{++}\left( 1-f\right) }-2\right]
\end{equation}
for $f$ around $f=u^{+i}u_{i}^{-}$ . Inserting these expressions (4.2, 4)
into the bosonic action (3.27), one obtains the following result:

\begin{equation}
\left.
\begin{array}{c}
S=\int d^{2}zdu\left\{ -\frac{12}{\lambda ^{2}}\left[ \frac{\partial
_{++}f^{++}\partial _{--}f^{++}}{f^{++^{2}}}\right] -\frac{4}{\lambda ^{2}}%
\left[ \frac{\partial _{++}f\partial _{--}f^{++}}{f^{++}\left( 1-f\right) }+%
\frac{\partial _{++}f^{++}\partial _{--}f}{\left( 1-f\right) f^{++}}\right] +%
\frac{4}{\lambda ^{2}}\left[ \frac{\partial _{++}f\partial _{--}f}{\left(
1-f\right) ^{2}}\right] \right. \\
+\frac{8}{\lambda ^{2}}\left[ \frac{\partial _{--}\partial _{++}f^{++}}{%
f^{++}}+\frac{\partial _{--}\partial _{++}f}{1-f}\right] -\frac{4\eta _{++}}{%
\lambda }\left[ \frac{f^{++}\partial _{--}f}{\xi ^{++^{2}}\left( 1-f\right)
^{2}}+\frac{f^{++^{2}}\partial _{--}f}{\xi ^{++^{2}}\left( 1-f\right) ^{3}}%
\right] \\
\left. -\frac{2}{\lambda }\left( \partial _{--}\eta _{++}+\partial _{++}\eta
_{--}-\alpha \overline{Z}\right) \left[ \frac{f^{++^{2}}}{\xi
^{++^{2}}\left( 1-f\right) ^{2}}\right] -\frac{\left| Z\right| ^{2}}{\lambda
^{2}}\left[ \frac{f^{++}}{\xi ^{++}\left( 1-f\right) }\right] \right\}
\end{array}
\right.
\end{equation}
This action contains terms with a singularity around
$f=u^{+i}u_{i}^{-}$. We point out that this singularity is shown
to originate from the solution of the Toda -(Liouville)- like
equation of motion (3.10). As it is worth stressing that not every
solution of the continual Toda-(Liouville)- equation of motion has
a good space-time interpretation, since the corresponding metric
might be incomplete with singularities, it seems at first sight
that our Hyper-Kahler metric will be incomplete. But using some
algebraic manipulations, one can derive the complete form of the
metric inspired from the first leading terms in a nice way.

To derive the metric from the bosonic action (4.5) , one carries out the
following operations step-by-step. Considering first the following
approximation:

\begin{equation}
\frac{1}{\left( 1-f\right) ^{\epsilon }}=\stackrel{\infty }{\stackunder{i=0}{%
\sum }}\frac{\left( \varepsilon +i-1\right) !}{\left( \epsilon -i\right) !i!}%
\text{ },\text{ }\varepsilon =1,2,3,...
\end{equation}
with $f=u_{(i}^{+}u_{j)}^{-}f^{\left( ij\right) }+f^{0}.$ Next, substitute
this expression into (4.5) and integrate over the harmonics once the power $%
f^{i}\left( Z,\overline{Z}\right) $ of the bosonic field $f$ in (4.6) are
expressed as series in terms of $f^{\left( ij\right) },$ $f^{0}$ and the
symmetrized product of harmonics. To do this one also has to use the
standard reduction identities [2]

\begin{equation}
u_{i}^{+}u_{(j_{1}...}^{+}u_{j_{n}}^{+}u_{k_{1}...}^{-}u_{k_{m})}^{-}=u_{(i}^{+}u_{j_{1}...}^{+}u_{j_{n}}^{+}u_{k_{1}...}^{-}u_{k_{m})}^{-}+%
\frac{m}{m+n+1}\varepsilon
_{i(k_{1}}u_{j_{1}...}^{+}u_{j_{n}}^{+}u_{k_{1}...}^{-}u_{k_{m})}^{-}
\end{equation}

\begin{equation}
u_{i}^{-}u_{(j_{1}...}^{+}u_{j_{n}}^{+}u_{k_{1}...}^{-}u_{k_{m})}^{-}=u_{(i}^{+}u_{j_{1}...}^{+}u_{j_{n}}^{+}u_{k_{1}...}^{-}u_{k_{m})}^{-}-%
\frac{m}{m+n+1}\varepsilon
_{i(j_{1}}u_{j_{1}...}^{+}u_{j_{n}}^{+}u_{k_{1}...}^{-}u_{k_{m})}^{-}
\end{equation}
and the $u_{i}^{\pm }$ integration rules.

\begin{equation}
\int du\left( u^{+}\right) ^{(m}\left( u^{-}\right) ^{n)}\left( u^{+}\right)
_{(k}\left( u^{-}\right) _{l)}=\left\{
\begin{array}{c}
\frac{\left( -1\right) ^{n}m!n!}{\left( m+n+1\right) !}\delta
_{(j_{1}}^{(i_{1}}...\delta _{j_{k+l})}^{i_{m+n})}\text{ \ }if\text{ }m=l,%
\text{ }n=k \\
0\text{ \ \ \ \ \ \ \ \ \ \ \ \ \ \ \ \ \ \ \ \ \ \ \ }otherwise
\end{array}
\right.
\end{equation}
with

\begin{equation}
\left( u^{+}\right) ^{(m}\left( u^{-}\right) ^{n)}\equiv
u^{+(i_{1}}...u^{+i_{m}}u^{-j_{1}}...u^{-j_{n})}
\end{equation}
Lengthy and very hard calculations lead finally to the following purely
bosonic action:

\begin{equation}
\left.
\begin{array}{c}
S\left[ f\right] =\int d^{2}z\left\{ A_{ijkl}\partial _{++}f^{\left(
ij\right) }\partial _{--}f^{\left( kl\right) }+B_{ij}\left[ \partial
_{++}f^{\left( ij\right) }\partial _{--}f^{0}+\partial _{++}f^{0}\partial
_{--}f^{\left( ij\right) }\right] \right. \\
\left. +C_{ij}\partial _{++}\partial _{--}f^{\left( ij\right) }+D\partial
_{++}\partial _{--}f^{0}+E\partial _{++}f^{0}\partial _{--}f^{0}+F\partial
_{++}f^{\left( ij\right) }\partial _{--}f_{\left( ij\right) }+V\left(
f\right) \right\}
\end{array}
\right.
\end{equation}
where

\begin{equation}
V\left( f\right) =G_{++}\partial _{--}f^{0}+K_{++\left( ij\right) }\partial
_{--}f^{\left( ij\right) }+L.
\end{equation}
From (4.10) one can easily derive the metric. The tensor components of this
metric are $A_{ijkl},$ $B_{ij},$ $C_{ij},$ $D,$ $E$ and $F,$ such that

\begin{equation}
\left.
\begin{array}{c}
A_{ijkl}=A_{jikl}=A_{ijlk} \\
B_{ij}=B_{ji} \\
C_{ij}=C_{ji}
\end{array}
\right.
\end{equation}
The first explicit expression obtained for this metric is of course
incomplete due to the previous approximation. The missing terms in this
metric as well as (the potential V(f)) are easily recuperated by looking
just at the behavior of the first leading terms of the components $A_{ijkl},$
$B_{ij},$ $C_{ij},$ $D,$ $E,$ $F$ and $\left( G_{++},\text{ }K_{++(ij)},%
\text{ }L\right) $. The complete expression of the metric described by the
bosonic field f reads as:

\begin{equation}
A_{ijkl}=-\frac{24}{\lambda ^{2}}\frac{1}{f^{\left( ij\right) }f^{\left(
kl\right) }}+\frac{2}{3\lambda ^{2}}\stackrel{}{\stackunder{N=0}{\sum }}%
\left\{ C_{N+1}^{N}f^{0N}+\stackrel{}{\stackunder{n=1}{\sum }}A\left(
N,n\right) f^{0N-2n}\left( ff\right) ^{n}\right\} \left( \frac{f_{\left(
ij\right) }}{f_{\left( kl\right) }}+\frac{f_{\left( kl\right) }}{f_{\left(
ij\right) }}\right)
\end{equation}

\begin{equation}
\left.
\begin{array}{c}
B_{ij}=-\frac{4}{3\lambda ^{2}}\stackrel{}{\stackunder{N=0}{\sum }}\left\{
2C_{N+2}^{N}f^{0N}+\stackrel{}{\stackunder{n=1}{\sum }}B_{1}\left(
N,n\right) f^{0N-2n}\left( ff\right) ^{n}\right\} f_{\left( ij\right) } \\
-\frac{8}{\lambda ^{2}}\stackrel{}{\stackunder{N=0}{\sum }}\left\{ f^{0N}+%
\stackrel{}{\stackunder{n=1}{\sum }}B_{2}\left( N,n\right) f^{0N-2n}\left(
ff\right) ^{n}\right\} \frac{1}{f^{\left( ij\right) }}
\end{array}
\right.
\end{equation}

\begin{equation}
C_{ij}=\frac{8}{\lambda ^{2}}\frac{1}{f^{\left( ij\right) }}-\frac{4}{%
3\lambda ^{2}}\stackrel{}{\stackunder{N=0}{\sum }}\left\{ C_{N+1}^{N}f^{0N}+%
\stackrel{}{\stackunder{n=1}{\sum }}C\left( N,n\right) f^{0N-2n}\left(
ff\right) ^{n}\right\} f_{\left( ij\right) }
\end{equation}

\begin{equation}
D=\frac{8}{\lambda ^{2}}\stackrel{}{\stackunder{N=0}{\sum }}\left\{ f^{0N}+%
\stackrel{}{\stackunder{n=1}{\sum }}D\left( N,n\right) f^{0N-2n}\left(
ff\right) ^{n}\right\}
\end{equation}

\begin{equation}
E=\frac{4}{\lambda ^{2}}\stackrel{}{\stackunder{N=0}{\sum }}\left\{
C_{N+1}^{N}f^{0N}+\stackrel{}{\stackunder{n=1}{\sum }}E\left( N,n\right)
f^{0N-2n}\left( ff\right) ^{n}\right\}
\end{equation}

\begin{equation}
F=-\frac{2}{3\lambda ^{2}}\stackrel{}{\stackunder{N=0}{\sum }}\left\{
C_{N+1}^{N}f^{0N}+\stackrel{}{\stackunder{n=1}{\sum }}F\left( N,n\right)
f^{0N-2n}\left( ff\right) ^{n}\right\}
\end{equation}
where $\left( ff\right) =f^{\left( ij\right) }f_{\left( ij\right) }.$

Furthermore the coefficients A(N,n), Bi(N,n),$\ldots $ , F(N,n) are finite
numerical quantities defined for $n\geq 1$ and $N\geq n$. Some examples are

\begin{equation}
\left.
\begin{array}{c}
A_{1}\left( 2,1\right) =-\frac{1}{5},\text{ }A_{2}\left( 2,1\right) =-\frac{%
15}{14},... \\
B_{1}\left( 2,1\right) =-\frac{4}{5},\text{ }B_{2}\left( 2,1\right) =-\frac{1%
}{6},... \\
C\left( 2,1\right) =-\frac{1}{5},\text{ }D\left( 2,1\right) =-\frac{1}{6},...
\\
E\left( 2,1\right) =-\frac{1}{2},\text{ }F\left( 2,1\right) =-\frac{1}{2},...%
\text{ }
\end{array}
\right.
\end{equation}
Another interesting consequence of the presence of central charges is a non
trivial scalar potential V(f) which is given by

\begin{equation}
\left.
\begin{array}{c}
V\left( f\right) =-\frac{8}{\lambda }\eta _{++}\stackrel{}{\stackunder{N=0}{%
\sum }}\left\{ C_{N+2}^{N}f^{0N}+\stackrel{}{\stackunder{n=1}{\sum }}G\left(
N,n\right) f^{0N-2n}\left( ff\right) ^{n}\right\} \frac{f^{\left( ij\right)
}f^{\left( kl\right) }}{\xi ^{\left( ij\right) }\xi ^{\left( kl\right) }}%
\partial _{--}f^{0} \\
+\frac{4}{3\lambda }\eta _{++}\stackrel{}{\stackunder{N=0}{\sum }}\left\{
3C_{N+3}^{N}f^{0N}+\stackrel{}{\stackunder{n=1}{\sum }}K\left( N,n\right)
f^{0N-2n}\left( ff\right) ^{n}\right\} \frac{f^{\left( kl\right) }f^{\left(
pq\right) }}{\xi ^{\left( kl\right) }\xi ^{\left( pq\right) }}f_{\left(
ij\right) }\partial _{--}f^{\left( ij\right) } \\
-\frac{4}{\lambda }\left( \partial _{--}\eta _{++}+\partial _{++}\eta
_{--}-\alpha \overline{Z}\right) \stackrel{}{\stackunder{N=0}{\sum }}\left\{
C_{N+1}^{N}f^{0N}+\stackrel{}{\stackunder{n=1}{\sum }}L_{1}\left( N,n\right)
f^{0N-2n}\left( ff\right) ^{n}\right\} \frac{f^{\left( ij\right) }f^{\left(
kl\right) }}{\xi ^{\left( ij\right) }\xi ^{\left( kl\right) }} \\
-\frac{\left| Z\right| ^{2}}{\lambda ^{2}}\stackrel{}{\stackunder{N=0}{\sum }%
}\left\{ f^{0N}+\stackrel{}{\stackunder{n=1}{\sum }}L_{2}\left( N,n\right)
f^{0N-2n}\left( ff\right) ^{n}\right\} \frac{f^{\left( ij\right) }}{\xi
^{\left( ij\right) }}
\end{array}
\right.
\end{equation}
where $\alpha $ and $\eta _{rr}$ are arbitrary constant already introduced
in section 3. The coefficients G(N,n), K(N,n) and Li(N,n) are also finite
numerical quantities defined for $n\geq 1$ and $N\geq 2n$. Particular
examples are given by

\begin{equation}
\left.
\begin{array}{c}
G\left( 2,1\right) =-1,\text{ }G\left( 3,1\right) =-5,\text{ }K\left(
2,1\right) =-\frac{1}{3},\text{ }K\left( 3,1\right) =-2, \\
L_{1}\left( 2,1\right) =-\frac{1}{2},\text{ }L_{1}\left( 3,1\right) =-2,%
\text{ }L_{2}\left( 2,1\right) =-\frac{1}{6},\text{ }L_{2}\left( 3,1\right)
=-\frac{1}{2},...
\end{array}
\right.
\end{equation}
Note that a possibility of generating non-trivial scalar potential via
non-vanishing central charges in the non linear 2D, N=4 supersymmetric
sigma-models was noticed earlier by Alvarez-Gaum\'{e} and Freedman in ref
[5].

\section{Conclusion and discussion:}

In the spirit to extend the results already established in $\left[ 3\right] $%
and which deal with an explicit derivation of a new hyperkahler
metric associated to D=2 N=4 SU(2) Liouville self interacting
model, we have tried to look for the contribution of central
charges operators in the theory under consideration. This central
extension is shown to give rise to a non trivial scalar potential
$V\left( f\right) $ depending on the physical bosonic field $f$.
The major content of this work can be summarized as follows:

1. Having shown how the Liouville-like equation of motion
Eq$\left( 3.5\right) $ leads to the bosonic central extended
action Eq$\left( 3.17\right) $ depending on $\omega ,B_{\pm \pm
}^{--},\overline{F}^{--}$, our first principal task was to search
to reduce much more this action, a fact which will help to extract
easily the associated hyperkahler metric. To do this one have to
solve the non linear differential equations for the auxiliary
fields $B_{\pm \pm }^{--},\overline{F}^{--}.$ Typical solutions
generalizing the ones obtained in $\left[ 3\right] $ are proposed.
The originality related to these equations \ consists in solving
not only the non
linear equations for the fields $B_{\pm \pm }^{--}$ but also for the field $%
\overline{F}^{--}$ (recall that the latter is integrated out when $Z=%
\overline{Z}=0$ as shown in $\left[ 3\right] $ ). The proposed
explicit solutions are given in Eqs(3.18-19) and the resulting
reduced bosonic action is shown to take the form presented in
Eq(3.27).

2. In computing the metric, we have observe that the factor proportional to $%
\left| Z\right| ^{2}$ in the action Eq(3.27) namely $\omega ^{2}+\frac{2}{%
\lambda ^{2}}$ is nothing but the first leading contribution of the non
linear function $\frac{e^{\lambda \omega }+e^{-\lambda \omega }}{\lambda ^{2}%
}$ for which $e^{\lambda \omega }$ is just the Liouville scalar
potential satisfying Eq(3.10). This is an important step towards
achieving the reduction procedure of the action to the purely
bosonic form. One of the advantages of this observation is that
one can overcome the difficulty in treating the term $\omega
^{2}+\frac{2}{\lambda ^{2}}$ by using simply Eq(4.1) which gives
approximately a factor $\frac{\left| Z\right| ^{2}}{ \lambda
^{2}}\left( \frac{f^{++}}{\xi ^{++}\left( 1-f\right) }+\frac{\xi
^{++}\left( 1-f\right) }{f^{++}}\right) .$

3. The use of the approximation Eq$\left( 4.6\right) $ is
important in computing the singular term in Eq$\left( 4.5\right) $
once the standard reduction identities are used. The missing terms
in doing the algebraic computations are naturally recuperated by
looking just at the behaviour of the first leading ones. This
gives once again the possibility of interpreting the resulting
metric associated to the $\left( Z,\overline{Z}\right) $ extended
D=2 N=4 SU(2) Liouville self interacting model as a complete
hyperkahler metric.

4. Another interesting consequence of introducing central charges
in the action is the possibility to derive explicitly a non
trivial scalar potential V(f) which shows among other a dependence
in $\left| Z\right| ^{2}$. The component form on this scalar
potential is given in Eq$\left(4.20\right) $. Setting for example
the arbitrary constant $\alpha $ and $ \eta _{rr}$ to be zero, one
obtain the following expression for the scalar potential

\[
V\left( f\right) =-\frac{\left| Z\right| ^{2}}{\lambda ^{2}}\stackrel{}{%
\stackunder{N=0}{\sum }}\left\{ f^{0N}+\stackrel{}{\stackunder{n=1}{\sum }}%
M_{2}\left( N,n\right) f^{0N-2n}\left( ff\right) ^{n}\right\} \frac{%
f^{\left( ij\right) }}{\xi ^{\left( ij\right) }}
\]
Moreover its important to note that our potential shre with the induced D=2
N=4 Taub-Nut hypermultiplet self interacting potential $\left[ 17\right] $
namely

\[
V\left( f\right) =\left| Z\right| ^{2}\frac{f\overline{f}}{1+\lambda f%
\overline{f}},
\]
the dependence in $\left| Z\right| ^{2}$, a term which we can also interpret
as a BPS-like mass operator.

5. Once the two integration over the Grassman variables and the harmonics
are done, the action associated to the $\left( Z,\overline{Z}\right) $
extended D=2 N=4 SU(2) Liouville self interacting model is shown to takes
the form given in Eq(4.10). From this action one can expliitly derive the
metric given by the tensors components explicitly presented in Eqs(4.12-18).
An important remarque conserning this hyperkahler metric is that in each
term of the components $A_{ijkl},B_{ij},C_{ij},D,E$ $and$ $F$, we have a
general coefficient given by

\[
K\left( \ast ,N,n,f\right) =\left( \ast \right) f^{0N}+\stackrel{}{%
\stackunder{n=1}{\sum }}\left( \ast \right) f^{0N-2n}\left( ff\right) ^{n}
\]
a property which is also shown in the derived potential $V\left( f\right) $.

6. It is now widely known that harmonic superspace provides a
framework for constructing general hyperkahler metrics. Original
results in this issue are given by the Taub-Nut and Egushi-Hanson
metrics exhibiting both U(1) Pauli-Gursey isomertie as shown by
Gibbon at all in $\left[ 11-c\right] $. In parallel to this work,
the authors showed in an original way, the existence of the
multicenter family of hyperkahler metrics including the above ones
and their integrable deformations $\left[ 11-b,c\right] $. They
showed also that these metrics are associated to a series of
potential depending explicitly on the harmonics variables. As our
D=2 N=4 SU(2) Liouville like potential Eq (2.14) share with the
Egushi-Hanson potential \ the impotrant feature to incorporaete
both the dimensionless quantity $\xi ^{++}$ which shows an
explicit dependence on the harmonics, we can then conclude that
our derived hyperkahler metric belongs to the Gibbons et al.
multicenter family exhibiting a U(1) symmetry and breaking SU(2)
invariance.

\section{\bf References}

\begin{enumerate}
\item[{[1]}] A.Galperin, E.Ivanov, V. Ogievetsky and E. Sokatchev, JETP,
Lett.40(1984)912; A.Galperin, E.Ivanov, S.Lalitzin, V. Ogievetsky
and E.Sokatchev, Class.Quant.Grav.1 (1984) 469.

\item[{[2]}] A.Galperin,
E.Ivanov, V.Ogievetsky and P.K. Townsend, Class.Quant.Grav.3
(1984) 625; A. Galperin,E.Ivanov, V.Ogievetsky and E. Sokatchev,
Comm.Math.Phys.103 (1986) 515.

\item[{[3]}] M.B.Sedra, Nucl.Phys.B 513 (1998) 709-722.

\item[{[4]}] M. Atiyah and N.Hitchin, phys.Lett.A 107 (1985) 21; G.Gibbons
and N.Manton, Nucl phys.B 274 (1986) 183.

\item[{[5]}] L. Alvarez Gaume and D.Z.Freedman, Comm.Math.Phys.80 (1981)
443.

\item[{[6]}] N.J. Hitchin, A. Karlhede, U.Lindstrom and M.Rocek,
Comm.Math.Phys.108 (1987) 535.

\item[{[7]}] G.W.Gibbons and N.S.Manton, Phys.Lett.B 356 (1995) 32;
G.W.Gibbons and P.Rychenkova, hep-th/ 9608085.

\item[{[8]}] E.H.Saidi and M.B.Sedra, Int. Jour. Mod. Phys.
AV9N6(1994)891-913.

\item[{[9]}] E.H.Saidi and M.B.Sedra, Mod.Phys.Lett.A 9 (1994) 3163.

\item[{[10]}] P.How, K.Stelle and P.K. Townsend, Nucl.Phys.B 214 (1983)
319.

\item[{[11]}] A.Galperin,E.Ivanov,V.Ogievetsky, Nucl.Phys.B 214 (1987) 74.

\item[{[12]}] U.U.Lindstrom and M.Rocek, Nucl.Phys.B 222 (1983)285.

\item[{[13]}] G.Gibbons and S.Hawking, Comm.Math.Phys.66 (1979)291;
G.Gibbons and P.Rubback , Comm.Math.Phys.115 (1988) 267;
G.Gibbons, D. Olivier, P.Rubback and G. Valent, Nucl.Phys.B
296(1988)679-696.

\item[{[14]}] A.Galperin, E.Ivanov, V.Ogievetsky and Sokatchev,
Comm.Math.Phys.103(1986) 515.

\item[{[15]}] T.Eguchi, P.Gilkey and A.Hanson, phys. Rep.66 (1980) 213.

\item[{[16]}] A. Leznov and M.Saviliev, Lett. Math.Phys.3 (1979)489;
Comm.Math.Phys.74(1980) 111; P.Mansfield , Nucl.Phys.B 208 (1982)
277.

\item[{[17]}] E.A.Ivanov, S.V.Ketov and B.M.Zupnik hep-th/ 9706078. [18]
A.Hanany and E.Witten, Princeton preprint IASSNS-HEP-96-121;
hep-th/ 9611230.\

\item[{[18]}] E.Witten, Princeton preprint IASSNS-HEP-97-19;
hep-th/9703166.\
\end{enumerate}
\end{document}